\documentclass[aps,pra,twocolumn,showpacs,superscriptaddress,showkeys]{revtex4-1}

\usepackage[utf8]{inputenc}
\usepackage[T1]{fontenc}

\usepackage{multirow}	
\usepackage{enumerate}	
\usepackage{Tabbing}	
\usepackage{relsize}	
\usepackage{textcomp}	
\usepackage{soul}	

\usepackage{amsmath}
\usepackage{amssymb}
\usepackage{bm}		

\usepackage{hyperref}
\usepackage{url}
\usepackage{cleveref}	

\usepackage{dcolumn}	
\usepackage{longtable}	
\usepackage{booktabs}	

\usepackage[pdftex]{graphicx}
\usepackage{epstopdf}	
\usepackage{rotating}	
\usepackage{float}	

\usepackage{listings}
\usepackage{color}
\usepackage{textcomp}

\definecolor{listinggray}{gray}{0.9}
\definecolor{lbcolor}{rgb}{0.9,0.9,0.9}

\usepackage{color}

\newcommand{\ket}[1]{\ensuremath{\left|#1\right\rangle}}
\newcommand{\average}[1]{\ensuremath{\left\langle#1\right\rangle}}
\newcommand{\bracket}[2]{\ensuremath{\left\langle#1 \vphantom{#2}\right| \left. #2 \vphantom{#1}\right\rangle}}
\newcommand{\matrixel}[3]{\ensuremath{\left\langle #1 \vphantom{#2#3} \right| #2 \left| #3 \vphantom{#1#2} \right\rangle}}

\newcommand{\AN}[2]{
\ensuremath{\hat{c}^{} _{{#1}
\ifnum#2=1  \uparrow
\else
\ifnum#2=-1  \downarrow
\else
\ifnum#2=2  \sigma
\else
\ifnum#2=-2  \bar{\sigma}
\else
\ifnum#2=3  \sigma '
\fi
\fi
\fi
\fi
\fi
}}
}
\newcommand{\CR}[2]{
\ensuremath{\hat{c}^\dagger _{{#1}
\ifnum#2=1  \uparrow
\else
\ifnum#2=-1  \downarrow
\else
\ifnum#2=2  \sigma
\else
\ifnum#2=-2  \bar{\sigma}
\else
\ifnum#2=3  \sigma '
\fi
\fi
\fi
\fi
\fi
}}
}
\newcommand{\NUM}[2]{
\ensuremath{\hat{n}^{} _{{#1}
\ifnum#2=1  \uparrow
\else
\ifnum#2=-1  \downarrow
\else
\ifnum#2=2  \sigma
\else
\ifnum#2=-2  \bar{\sigma}
\else
\ifnum#2=3  \sigma '
\fi
\fi
\fi
\fi
\fi
}}
}

\newcommand{\mc}[3]{\multicolumn{#1}{#2}{#3}}

\renewcommand{\vec}[1]{\ensuremath{\mathbf{#1}}}

\newcommand{\ComInEq}[2]{\ensuremath{\stackrel{#2}{\overbrace{#1}}}}

\definecolor{newRed}{RGB}{200,0,0}

\definecolor{newGreen}{RGB}{0,100,0}

\newcommand{\none}{\ensuremath{\varnothing}}

\begin{document}

\title{ Combined shared and distributed memory \emph{ab-initio} computations of molecular-hydrogen systems in the correlated state: process pool solution and two-level parallelism}

\author{Andrzej Biborski}
\email {andrzej.biborski@agh.edu.pl}
\affiliation {Academic Centre for Materials and Nanotechnology,
AGH University of Science and Technology,
al. A. Mickiewicza 30,  30-059 Krakow,  Poland}
 
\author{Andrzej P. K\k{a}dzielawa}
\email{kadzielawa@th.if.uj.edu.pl}
\affiliation{Marian Smoluchowski Institute of Physics, Jagiellonian University, 
ulica \L{}ojasiewicza 11, PL-30-348 Krak\'ow, Poland}

\author{J\'{o}zef Spa\l{}ek}
\email{ufspalek@if.uj.edu.pl}
\affiliation{Marian Smoluchowski Institute of Physics, Jagiellonian University, 
ulica \L{}ojasiewicza 11, PL-30-348 Krak\'ow, Poland}
\affiliation {Academic Centre for Materials and Nanotechnology,
AGH University of Science and Technology,
al. A. Mickiewicza 30,  30-059 Krakow,  Poland}

\date{\today}

\begin{abstract}
An efficient computational scheme devised for investigations of ground state properties of the electronically correlated systems is presented.
As an example,  $(H_{2})_{n}$  chain  is considered with
the long-range electron-electron interactions taken into account. 
The implemented procedure covers: (\textit{i}) single-particle Wannier wave-function basis construction in the correlated state, (\textit{ii}) microscopic parameters calculation, and
(\textit{iii}) ground state energy optimization.
The optimization loop is based on highly effective \emph{process-pool} solution -- specific \emph{root--workers} approach. The hierarchical, two-level parallelism  was applied: both shared (by use of Open Multi-Processing) and distributed (by use of Message Passing Interface) memory models were utilized.
We discuss in detail the feature that such approach results in a substantial increase of the calculation speed reaching   factor of $300$  for the fully parallelized solution.
The elaborated in detail scheme reflects the situation in which the most demanding task is the single-particle basis optimization.
\end{abstract}


\maketitle

\section{Physical Motivation: Exact Diagonalization + ab Initio Method}
\label{sec:motivation}

Electronically correlated systems are important both from the point of view of their unique physical properties  and from nontrivial 
computational methods developed to determine them.
The latter cover  methods based on the Density Functional Theory (DFT) with the energy functional enriched by the  correlation terms
-- the on-site repulsion $U$ in the Hubbard model \cite{DFTU} and the Hund's rule term in the case of orbital degeneracy. Often, they are incorporated 
into either DFT or the Dynamic Mean Field Theory (DMFT) approach supplemented with the LDA-type calculations (see e.g. \cite{Karolak}).
On the other hand, the \emph{Configuration-Interaction} (CI) method  does not suffer from 
the well--known \emph{double counting problem}\cite{DFTU,Karolak}, inherent in the DFT+U or LDA+DMFT methods.
Another approach, similar in its spirit to the CI method, formulated as a combination of \emph{the first-} and \emph{second quantization} (FQ, SQ respectively) 
formalisms  was elaborated in our group in the last decade and termed the \textbf{E}xact \textbf{D}iagonalization \textbf{Ab} \textbf{I}ntito (EDABI) approach \cite{Spalek1,Spalek2}.
This method allows for a natural incorporation of the correlation effects consistently by the advantages of using the SQ language so that the
\emph{double-counting problem} does not arise at all. Also, by construction, it includes the Pauli principle for the fermionic systems.
In contrast to CI the EDABI approach avoids any direct dealing with the many--body wave function expressed via a linear combination of the  Slater determinants \cite{Surjan}.
Instead, it is based  on the  many--particle quantum states constructed in the occupation number representation \cite{Surjan} -- standard procedure for the SQ formulated problems.

The  application of EDABI  was found promising in view of research devoted to the hydrogen molecular systems with inclusion of interelectronic correlations \cite{Kadzielawa2},
nano-clusters \cite{RycerzPhD}, and to atomic hydrogen metallization \cite{Kadzielawa1}.
As the many--particle state is explicitly written in the occupation-number representation (Fock space), the starting Hamiltonian is formulated in the SQ language.
Electronic correlations are then automatically  included
in the modeled system. However,  in the Hubbard--like starting Hamiltonians \cite{Hubbard1,Gutzwiller1,Gutzwiller2,Kanamori} the knowledge of microscopic parameters, 
such as the on-site energy, the intersite hoppings, and the Coulomb repulsion magnitudes are regarded as input information.
These parameters are often estimated indirectly. With this limitation, specific \emph{phase-diagrams} are  constructed and the 
phase boundaries of interest are determined as a function of those microscopic parameters which are not directly measurable (cf. e.g. \cite{Zegrodnik1, Zegrodnik2,WysokinskiAbram}). 
In EDABI we take a different route: relatively simple and small systems are to be described consistently in the sense that the microscopic parameters are
obtained explicitly as an output of an appropriate ab-initio variational procedure. 
Therefore, the EDABI approach should be regarded as an \emph{ab--initio} method but with the single-particle wave functions being determined self-consistently in
the correlated state.
In this manner, the problem solution is reversed with respect to that in either LDA+U or LDA+DMFT. Namely, we first formulate
the Hamiltonian and diagonalize it in SQ formalism and determine the single-particle wave function only as a second step.
However  realistic, such an approach to the electronically correlated systems implies a substantially greater 
computational complexity, since the variational optimization  consists of  (\emph{i}) microscopic parameters calculation and (\emph{ii}) concomitant Hamiltonian-matrix diagonalization.
We address here the issue (\emph{i}) presenting how the modern High Performance Computing  (HPC) cluster architecture can be utilized in the context, where the number of microscopic
parameters is substantial if not large,
and their calculation is one of the  potential \emph{bottlenecks} in the whole computational procedure.
We also provide an example how  a many-body  problem at hand may be supplemented with the two-level parallelism in an intuitive manner. We do not discuss 
either the methodology related to the point (\emph{ii}) or to its algorithmic aspect or else, to  technical opportunities provided by e.g. recent fast development of Graphic
Processing Units (GPU) computational techniques that are also in the area of interest \cite{Siro,Wendt}. Nonetheless, as it
becomes clear below, heterogeneous solutions are also easily applicable in our scheme.

The structure of the paper is as follows.
In Sec.~\ref{sec:EDABI} we describe briefly the EDABI method (cf. Appendix~\ref{app:EDABI} for details) and emphasize  the computational complexity aspects. Next, in Sec.~\ref{sec:parallel}, we show how the \emph{process--pool} concept enhanced by \emph{the two-level parallelism} forms a natural 
solution. In Sec.~\ref{sec:model} we present the outcome of its implementation: results of calculations carried out for $(H_{2})_{n}$ exemplary system and discuss the achieved speed--up when compared to the  reference single CPU  computations.
In Appendix~\ref{app:Convergence} we discuss the convergence of our model for the case of infinite systems.

\section{Computational method - EDABI}
\label{sec:EDABI}
As stated in the foregoing Section, the computational method considered here is based on the EDABI method, comprehensive description of which can be found e.g. in \cite{Spalek1}. It allows for consideration of
realistic, electronically--correlated nanosystems within the framework of the combined first-- and second-quantization formalisms. Below we sketch this method (for details see Appendix~\ref{app:EDABI}).

\subsection{Second-quantization aspect}
For the purpose of calculating the ground-state energy of given system we start with the second-quantization language \cite{Dirac,Fock,Surjan,FetterWalecka}. 
We introduce the fermonic \emph{anihilator}(\emph{creator}) $\hat{c}^{(\dagger)} _{i \sigma}$ algebra by imposing the anticommutation relations among them, namely
\begin{align}
 \{ \CR{i}{2}, \CR{j}{3} \} \equiv \{ \AN{i}{2}, \AN{j}{3} \} \equiv 0 \ \ \ \text{and} \ \ \{ \CR{i}{2}, \AN{j}{3} \} \equiv \delta_{ij}\delta_{\sigma \sigma'},
\end{align}
where $i$ and $j$ denote sites (nodes) of a fixed lattice, $\sigma, \ \sigma' = \pm 1$ are the spin quantum number, and the anticommutator is $\{ A, B \} \equiv AB + BA$. 

We represent the many-particle basis states $\{\ket{\Phi_k}\}$ on the lattice of $\Lambda$ sites in the \emph{Fock space} \cite{Fock} in the following manner 
\begin{align}
	\label{eq:basis_states}
	\ket{\Phi_k} =& \prod_{i \in \Omega_{\uparrow k}} \CR{i}{1} \prod_{j \in \Omega_{\downarrow k}} \CR{j}{1} \ket{0},
\end{align}
where $\Omega_{\uparrow k}$ and $\Omega_{\downarrow k}$ are the subsets of sites occupied by fermions with $\Lambda_\uparrow$ and  $\Lambda_\downarrow$ particles respectively, and $\ket{0}$ is
the \emph{vacuum} state (with no particles), with $\bracket{0}{0} \equiv 1$. Explicitly,
\begin{align}
\label{eq:basis_states_eg}
 \ket{\Phi} =& \underbrace{\ket{0, 1, \dots, 1}}_{\text{spin } \uparrow} \otimes \underbrace{\ket{1, 0, \dots, 1}}_{\text{spin } \downarrow} = \\\notag
	    =& \CR{2}{1} \cdots \CR{\Lambda}{1} \CR{1}{-1} \cdots \CR{\Lambda}{-1} \ket{0}.
\end{align}
With this concrete (occupation-number) representation of an abstract Fock space we define next the microscopic Hamiltonian of our interacting
system of fermions.

\subsection{Definition of the physical problem}
We take the real-space representation with the starting
field operators in the form
\begin{align}
 \label{eq:field_opers}
 \hat{\Psi}^{\phantom{\dagger}}_\sigma ( \vec{r} ) = \sum_{i} w_i ( \vec{r} ) \chi_\sigma \AN{i}{2},
\end{align}
where $w_i ( \vec{r} )$ is the single-particle wave function for fermion (e.g. electron) located on $i$-th site, $\chi_\sigma$ is the spin wave function
($\sigma = \pm 1$) with global spin quantization axis ($z$-axis). In general, the many-particle Hamiltonian is defined in the form
\begin{align}
 \label{eq:hamiltonian_general}
 \hat{\mathcal{H}} =& \sum_\sigma \int d^3r \hat{\Psi}^{{\dagger}}_\sigma  ( \vec{r}  ) \hat{\mathcal{H}}_1  ( \vec{r}  ) \hat{\Psi}^{\phantom{\dagger}}_\sigma  ( \vec{r}  ) \\\notag
		   &+ \frac{1}{2} \sum_{\sigma \sigma'} \iint d^3r d^3r' \hat{\Psi}^{{\dagger}}_\sigma  ( \vec{r}  ) \hat{\Psi}^{{\dagger}}_{\sigma'}  ( \vec{r}'  ) \hat{V}  ( \vec{r} - \vec{r}'  ) \hat{\Psi}^{\phantom{\dagger}}_{\sigma'}  ( \vec{r}'  ) \hat{\Psi}^{\phantom{\dagger}}_\sigma  ( \vec{r}  ),
\end{align}
where $\hat{\mathcal{H}}_1$ is the (spin-independent) Hamiltonian for a single particle in the milieu of all other particles and $\hat{V}  ( \vec{r} - \vec{r}'  )$ 
is the interaction energy for a single pair. For the modeling purposes we assume that $\hat{\mathcal{H}}_1$ is expressed in the atomic units ($\hbar = e^2 / 2 = 2 m_e = 1$, where $e$ is the charge of an electron and
$m_e$ is its mass) and expresses the particle kinetic energy and the attractive interaction with the protons located at $\{\vec{R}_i\}$, i.e.,
\begin{align}
 \label{eq:one-body-ham}
 \hat{\mathcal{H}}_1  ( \vec{r}  ) &\overset{a.u.}{=} -\nabla^{2} - \sum_{i=1}^{N_S}\frac{2}{|\vec{R_i} - \vec{r}|},
\end{align}
where $N_S$ is the number of sites, whereas 
\begin{align}
  \label{eq:two-body-ham}
 \hat{V}( \vec{r} - \vec{r}'  ) &\overset{a.u.}{=}  \frac{2}{|\vec{r} - \vec{r'}|},
\end{align}
represents the Coulomb repulsive interaction between them.
Substituting \eqref{eq:field_opers} into \eqref{eq:hamiltonian_general} we obtain the explicit second-quantized form of the Hamiltonian \cite{FetterWalecka,Surjan} i.e.,
\begin{align}
 \label{eq:hamiltonian}
  \mathcal{H} = \sum\limits_{ij}\sum\limits_{\sigma}t_{ij}\CR{i}{2}\AN{j}{2} +\sum\limits_{ijkl}\sum\limits_{\sigma,\sigma'}V_{ijkl}\CR{i}{2}\CR{j}{3}\AN{l}{3}\AN{k}{2},
\end{align}
where $t_{ij}$ and $V_{ijkl}$ are integrals associated with the  one- and two-body operators respectively
\begin{subequations}
\label{eq:microscopicP}
\begin{align}
 \label{eq:one-body}
  t_{ij} &\equiv \matrixel{w_{i}(\vec{r})}{\hat{\mathcal{H}}_1}{w_j(\vec{r})} \\\notag
         &= \int d^3r \ w_i^*(\vec{r}) \hat{\mathcal{H}}_1 (\vec{r}) w_j(\vec{r}) , \\
  \label{eq:two-body}
  V_{ijkl} &\equiv \matrixel{w_i(\vec{r}) w_j(\vec{r'})}{\hat{V}}{w_k(\vec{r})w_l(\vec{r}')} \\\notag
	   &= \iint d^3r d^3r' \ w_i^*(\vec{r}) w_j^*(\vec{r}') \hat{V}  ( \vec{r} - \vec{r}'  ) w_k(\vec{r})  w_l(\vec{r}') .
\end{align}
\end{subequations}
The first term contains the single-particle part composed of the atomic energy $\epsilon_i \equiv t_{ii}$, as well as
expresses the kinetic (hopping) part with $t_{ij}$ ($i\neq j$) being the so-called hopping integral.
The second expression contains intraatomic
(intrasite) part of the interaction between the particles ($ U_i \equiv V_{iiii}$ -- the so-called Hubbard interaction), and
the intersite (interatomic) interaction ($K_{ij} \equiv V_{ijij}$ -- the last important term for the purposes here).
A remark is in place here: when the single-particle basis $\{w_i\}$ is assumed as real, the last two two-body interaction terms are the exchange-correlation energy ($J_{ij} \equiv V_{ijji}$) and
the so-called correlated hopping ($V_{ij} \equiv V_{ijjj}$).

The Hamiltonian \eqref{eq:hamiltonian_general}, with inclusion of all the two-site terms only,
can be rewritten in the following form, with the microscopic parameters contained in an explicit manner, i.e.,
\begin{align}
 \label{eq:hamiltonian_special}
 \hat{\mathcal{H}} &= \sum_{i,\sigma}\epsilon_{i} \NUM{i\sigma}{0} + \frac{1}{2}\sum_{\sigma,i\neq j} t_{ij} \CR{i\sigma}{0} \AN{j\sigma}{0}+\frac{1}{2}\sum_{i,\sigma}U_{i}\NUM{i}{2} \NUM{i}{-2}\\\notag
&- \sum_{i \neq j} J_{ij} \vec{S}_i \cdotp \vec{S}_j + \frac{1}{2}\sum_{i\neq j}\left(K_{ij} - \frac{J_{ij}}{2}\right)\NUM{i}{0} \NUM{j}{0}\\\notag
& + \sum_{i \neq j}J_{ij} \CR{i}{1} \CR{i}{-1} \AN{j}{-1} \AN{j}{1} + \sum_{\sigma,i\neq j}V_{ij} \NUM{i}{2} \left( \CR{i}{-2}\AN{j}{-2} + \CR{j}{-2}\AN{i}{-2} \right).
\end{align}

The basis $\big\{w_i(\vec{r})\big\}_{\alpha}$ in \eqref{eq:field_opers} needs to be orthonormal, i.e. orthogonal and normalized to unity.
In the next Section, we describe how to  construct the basis satisfying this condition.
In summary, by solving (diagonalizing) $\hat{\mathcal{H}}$ we understand finding the optimal many-particle configuration
with a simultaneous single-particle basis $\{w_i(\vec{r})\}$ determination. Typically \cite{Zegrodnik1,Zegrodnik2,WysokinskiAbram,WysokinskiAbram1},
the parameters $\epsilon_i=\epsilon$, $t_{ij}$, $U$, $K_{ij}$ are regarded as extra parameters. Here we calculate them explicitly along with
the diagonalization in the Fock space at the same time.

\subsection{Basis orthogonalization as a bilinear problem} 
We require the orthonormality of the set of the single--particle wave functions $\left\{w_i(\vec{r}, \alpha)\right\}$, i.e., set the conditions
\begin{align}
 \label{eq:orthonormality}
	\bracket{w_i(\vec{r})}{w_j(\vec{r})} \equiv& \phantom{0} \\\notag &\int_{\mathbb{R}^3} d^3r \  w(\vec{r}-\vec{R}_i)w(\vec{r}-\vec{R}_j) = \delta_{ij},
\end{align}
where $\delta_{ij}$ is the Kronecker delta. Note that  $\alpha$ will play a role of variational parameter specifying the way
of constructing the basis (cf. Sec.~\ref{ssec:optimization}).
Namely, the single--particle wave functions (Wannier functions) are approximated by a finite linear combination in a selected set. These wave functions describe the single-electron states
centered on every atomic/ionic site, i.e., at positions $\big\{\vec{R_i}\big\}$. Such approach is related to the \emph{tight-binding}
approximation (TBA \cite{AshcroftMermin}), where the atomic orbitals composing $w_i$ are  represented by e.g.
the Slater-type orbitals (STO). For the purpose of the present model analysis, only the $1s$ Slater functions are taken into account, i.e.,
\begin{align}
\label{eq:slater}
	\psi_i \left(\vec{r}\right) \equiv \sqrt{\frac{\alpha^3}{\pi}} e^{-\alpha \left| \vec{r} - \vec{R}_i \right| },
\end{align}
where $\alpha$ is the inverse wave-function size. Similarly to \cite{Mulliken}, for each position $i$ we construct linear combination 
\begin{align}
 \label{eq:wannier_definition}
  w_i \left(\vec{r}\right) = \sum_{j=0}^{L}	\beta_j \widetilde{\psi}_{\pi_i \left( j \right)} \left(\vec{r} \right),
\end{align}
where $\{ \beta_j \}$ compose a set of $(L+1)$ mixing coefficients, and $\pi_i : \{0, \dots, L\} \rightarrow \mathcal{N}_i$,
is the function mapping indexes to the neighborhood $\mathcal{N}_i$ of the site (node) $i$ located at $\vec{R}_i$.

Note that in general $\widetilde{\psi}_{\pi_i \left( j \right)}$ may be a sum over Slater functions in the neighborhood, which varies
the number of $\beta$ coefficients and the number of nodes in the neighborhood $\mathcal{N}_i$. This circumstance does not influence the discussion, but is of crucial importance
when the scheme is implemented numerically.
Also, the new basis $\{ w_i (\vec{r}) \}$ is orthogonal \emph{in the neighborhood} $\mathcal{N}_i$.

We are looking for the set of $\{ \beta_j \}$, orthonormalizing the basis $\{ w_i (\vec{r}) \}$ for given geometry (effectively described
by set of ionic coordinates $\vec{R}_i$) and for the arbitrary inverse wave-function size $\alpha$. In order to achieve this we replace the original problem
$\forall j \in \big\{ \pi_i(k) \big| k \in \{ 0, 1, \dots, L \} \big\} $
\begin{align}
\int_{\mathbb{R}^3} d^3r \  w_i \left(\vec{r}\right)  w_j \left(\vec{r}\right) &= \delta_{ij} \\\notag
\end{align}
with the equivalent set of bilinear equations
\begin{align}
\label{eq:masterOrtho}
\underline{\beta}_i^T \mathbb{S}_{ij} \underline{\beta}_j &= \delta_{ij},
\end{align}
where
\begin{equation}
\underline{\beta}_i \equiv \left(
  \begin{array}{c}
  \beta_{\pi_i(0)}\\
  \beta_{\pi_i(1)}\\
  \vdots\\
  \beta_{\pi_i(L)}
  \end{array}
\right),
\end{equation}
and the overlap integrals are
\begin{align}
 \left( \mathbb{S}_{ij} \right) _{lm} \equiv  \int_{\mathbb{R}^3} d^3r \ \widetilde{\psi}_{\pi_i(l)}  \left(\vec{r}\right) \widetilde{\psi}_{\pi_j(m)} \left(\vec{r}\right),
\end{align}
with $\left( \mathbb{S}_{ii} \right) _{ll}=1$.
For given geometry and the inverse wave-function size $\alpha$ we solve the system \eqref{eq:masterOrtho} numerically.

The computation of the two-body integrals (cf. Eq.~\eqref{eq:two-body} ) must be performed in a general case numerically (see e.g. \cite{RycerzPhD} and citations therein). 
Therefore, STO are usually approximated by their expansion in the so-called Gaussian basis, namely
\begin{align}
 \label{eq:slater_definition}
 \psi_i \left(  \vec{r} \right) \approx  \alpha^{3/2}\sum_{a=1}^{N_G}\Big(  \frac{2\alpha^{2} \Gamma_{a}^{2}}{ \pi}\Big)^{3/4}e^{-\alpha^2\Gamma_{a}^2\vert\vec{R_{i}}-\vec{a}\vert^{2}},
\end{align}
with $N_G$ being the parameter describing number of Gaussian functions taken into account and the adjustable set  $\big\{\Gamma_{a}\big\}$ is obtained through
a separate procedure \cite{RycerzPhD}. 
Exact or approximate ground state properties (i.e. ground state energy, structural properties, electronic density, etc.) are 
obtained when the eigenstate corresponding 
to the  lowest many-particle eigenvalue is determined with the
diagonalization performed in the Fock space. In the context of EDABI, exact methods were successfully applied, e.g.,
the Lanczos \cite{Siro} algorithm for the matrix diagonalization, executable 
for nanosystems \cite{Spalek2, Kadzielawa2, RycerzPhD}. 

Although originally EDABI was formulated for finite-size systems, the scheme can be regarded as a general variational procedure. According to its generic character,
it is applicable in combination with another correlation oriented approach, dedicated to the approximate Hamiltonian diagonalization. As an example, the \emph{bulk} 
systems with  proper translational symmetries were
analyzed \cite{Kurzyk, Kadzielawa1}, based on the modified Gutzwiller Approximation (SGA).  One may incorporate other diagonalization schemes applicable to the EDABI method.
Here we consider only the scenario, according to which diagonalization in the Fock space
is performed \emph{exactly} -- i.e., the  Hamiltonian matrix is generated with the help of basis \eqref{eq:basis_states} and diagonalized in terms of iterative (e.g. Lanczos) numerical algorithm.

\subsection{Optimization procedure and computational complexity}
\label{ssec:optimization}
The set  $\left\{w_i(\vec{r},\alpha)\right\}$ describes the system in question: the interaction parameters
$\{V_{ijkl}\}$ and intersite hoppings $\{ t_{ij} \}$. On the other hand, there are independent parameters $\Big\{\alpha,\big\{\vec{R_i}\big\}\Big\}$,
where $\big\{\vec{R_i}\big\}$  together with $\alpha$ form the multidimensional optimization space. The EDABI method is based on the variational principle within which
the single-particle wave functions 
at given atomic configuration $\big\{\vec{R_i}\big\}$ are optimized to determine the ground-state energy of the correlated system. From the computational point of view, four main tasks are to be performed 
in a single iteration, i.e., for a given trial value of $\alpha$: 
\begin{enumerate}
 \item Single particle basis ortonormalization - solution of $L+1$ dimensional bilinear set of equations.
\label{en:01}
 \item Computation of the one-body microscopic parameters - scaling as $\mathcal{O}\big(L^2 N_G^2 N_S\big)$.
\label{en:02}
 \item Computation of the two-body microscopic parameters - scaling as $\mathcal{O}\big(L^{4}N_G^{4}\big)$.
\label{en:03}
 \item Hamiltonian diagonalization - dependent on the selected approach (exact, mean-field, Gutzwiller Approximation, etc.).
\label{en:04}
\end{enumerate}

The tasks corresponding to \ref{en:02} and \ref{en:03} are central to the subsequent considerations. While for relatively simple models, such as one band Hubbard model,
there are only three integrals to compute (the nearest-neighbor hopping,
the atomic reference site energy, and the onsite electrostatic repulsion), this is not the case in the situation, in which a more complicated 
Hamiltonian describes our system. The extended Hubbard model (see Sec.~\ref{sec:model}),
where the non-local electron-electron interactions are taken up to some cut-off distance, is associated with the increasing number of the two--body integrals to be computed.
However, also for the multiband Hubbard model case, 
the number of hopping integrals increases as $\mathcal{O}(N_{b}^2)$, where $N_b$ is the number of bands. Therefore, an effective scheme allowing to obtain -- possibly
quite large -- set of microscopic parameters in a run--time,
is desired. In the following Section we propose an explicit solution of this last issue.

\section{Process-pool Solution and Two-level Parallelism}
\label{sec:parallel}

As we said above, the standard task is to diagonalize Hamiltonian \eqref{eq:hamiltonian_general} defined in the Fock space (occupation-number representation). This means,
to determine the ground-state energy for given values of the microscopic parameters: $\epsilon_i$, $t_{ij}$, $U_i$, and $K_{ij}$ (in general, $J_{ij}$ and $V_{ij}$ as well).
The principal work we would like to undertake here is to determine the renormalized wave functions $\left\{ w_i (\vec{r}) \right\} _ {i=1,\dots,N}$ in the resultant
(correlated) ground state. The first aspect of the whole problem presents itself as an equally
important part, as only then the ground state configuration of our system can be defined physically, i.e., as a (periodic) system with known lattice parameter (interionic distance).
In the remaining part both aspects of the optimization problem are elaborated together with concomitant technical details provided.

\subsection{Optimization loop}
\label{ssec:optimization_loop}
We focus our analysis according to the scenario  that the computational
time spent on the diagonalization in the Fock space is negligible when compared to the calculation of the two-body integrals appearing in the calculation of microscopic
parameters. For the sake of clarity, let us rewrite Hamiltonian \eqref{eq:hamiltonian} in more compact form
 \begin{align}
 \mathcal{H} = \sum_m \sum_{ij} \Xi_{m;ij} \hat{O}_{m;ij},
\end{align}
where
$\Xi_{m;ij} \in \{ \epsilon_i, t_{ij}, U_i, K_{ij}, V_{ij}, J_{ij} \}$ and $\hat{O}_{m;ij}$ symbolizes the operator part,
e.g., $\hat{O}_{t;ij} = \sum_{\sigma} \CR{i}{2}\AN{j}{2}$.
One should note that the parameter set $\Xi_{m;ij}$ must be calculated in each iteration step during the optimization procedure. Computation of the 
microscopic parameters can be performed independently which in turn provides an opportunity for its acceleration by means of the parallelism application.

Let us consider some generic optimization procedure $OP(\big\{\vec{R_i}\big\})$ returning the minimal energy
at given accuracy, as a function of structural parameters. In  $OP$ the system energy is sampled as a function of $\alpha$ so we denote it as $E_G(\alpha)$. 
Taking into account that  $\nabla_{\alpha}E_G(\alpha)$ is 
not obtainable in general case, the optimization scheme encoded in $OP$ relates  to a non-gradient method, e.g., \emph{the golden--search} for 
the one-dimensional case. The computation of $E_G(\alpha)$ 
(sampled by $OP$) consists of calculation of $\{\Xi_{m;ij}\}$ combined with the Hamiltonian matrix diagonalization. Within our approach, the computation speed--up
is achieved  by implementing 
the  \emph{process-pool} or \emph{the root--worker processes}.
This solution might be regarded as \emph{a thread--pool} pattern, but constructed within the framework of the distributed memory model. Working threads are replaced by
the processes -- let us call them \emph{workers} -- which
may communicate by the utilizing the \emph{Message Passing Interface} (MPI) -- as it is done in our implementation.
\emph{Workers} remain in the infinite loop, monitoring signal from the \emph{root} process which in turn is responsible
for the job triggering and synchronization. It can also participate in the calculations -- in our case it performs matrix diagonalization.
Depending on what kind of signal is sent (in certain protocol established for the communication purposes
between \emph{workers} and  \emph{root}) workers: (\emph{i}) wait, (\emph{ii}) start to compute, (\emph{iii}) break and exit from the loop.
Since (\emph{ii}) might be considered as a generic task, one can see
that proposed approach is extendable to include diagonalization -- e.g., if one deals with the big block--diagonal matrices, each block could be
diagonalized independently by the worker processes. \emph{Process--pool} is supposed
to be efficient assuming that the following principal condition is fulfilled: the  task performed by each of the \emph{worker processes} (or at least by most of them) is
computationally the most expensive  part, particularly if it  shadows the communication latency. 

Each process in the \emph{process-pool} 
may utilize -- if there are available resources -- \emph{shared memory} model. Thus the solution benefits from \emph{two-level parallelism}, where \emph{worker--process} parent thread
forks into working threads, allowing in turn to perform each task  faster (integral computation in this context). In our case the elements from 
the set $\{\Xi_{m;ij}\}$ are distributed into the sub-sets $I_p = \{\Xi_{m;ij}\}_p$ where $p$ denotes the processor $id$. The $I_p$ relate to task stack assigned to $p$-th process. 
The sub-sets construction should be performed 
carefully to keep well-balanced workload on processor, e.g. providing equal distribution of the integrals calculation among $I_p$. The \emph{process-pool} consists 
of $P+1$ processes, $P$ of them are \emph{workers} and one -- as mentioned -- is a \emph{root}. As follows from the scheme (cf. Fig.~\ref{Fig:scheme}), \emph{process-pool} 
is applied to the EDABI optimization loop. The $OP$  procedure performs $\alpha$ space sampling along non-gradient optimization scheme. 
Each trial--$E_G$ computation demands parameters update and integrals calculations. The latter one exploits parallelism at two levels. 
Each of  $P$ processes computes 
integrals grouped in its assigned $I_p$ subset and each of the two--body integral  is calculated in the nested (fourth times) loops which are collapsed  
in terms of the  utilization of the openMP framework. The whole computation originates from the need to determine the two-body integrals
\eqref{eq:two-body} which are expressed as

\begin{align}
\label{eq:twobody_explicit}
  V_{ijkl} &= \langle w_i w_j|\mathcal{V}|w_k w_l \rangle = \\\notag
  &= \sum_{pqrs}\beta_{p}\beta_{q}\beta_{r}\beta_{s} \langle\widetilde{\psi}_{\pi_i \left( p \right)}\widetilde{\psi}_{\pi_j \left( q \right)}|\mathcal{V}|\widetilde{\psi}_{\pi_k \left( r \right)}\widetilde{\psi}_{\pi_l \left( s \right)}\rangle = \\\notag
&= \sum_{pqrs}V_{ijkl}^{pqrs},
\end{align}

 where $\widetilde{\psi}$ represents the Gaussian contraction. Elements $V_{ijkl}^{pqrs}$ are computable independently;
 therefore $V_{ijkl}$ is obtained with the help of openMP \emph{reduction} clause.  
Computed integrals are gathered in a single array in terms of $MPI\_Gather$ function or optionaly of $MPI\_Allgather$
if necessary (which potentially may increase communication latency). Then, Hamiltonian matrix is updated with the proper
values and the diagonalization step  starts. As an output of diagonalization,
the trial $E_G$ is computed and processed in $OP$.
Our implementation bases on MPI and openMP, though its scheme is generic and might be implemented by means of any of known technologies or self--made implementations as well.

\begin{widetext}
\begin{center}
\begin{figure}
 \includegraphics[width=.8\linewidth]{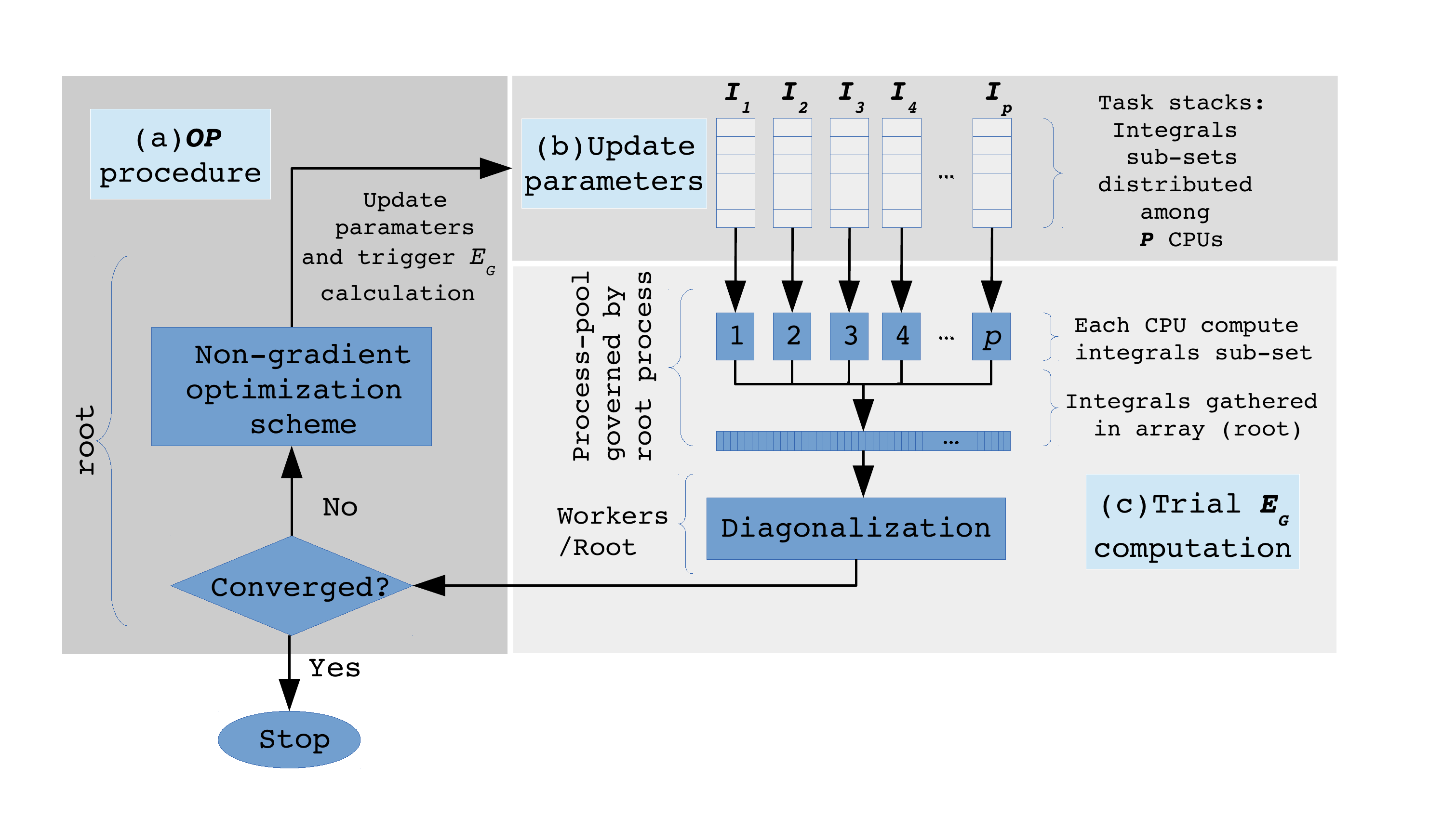}
 \caption{Process-pool solution applied for the optimization procedure ($OP$) in the context EDABI method.}
  \label{Fig:scheme}
\end{figure}
\end{center}
\end{widetext}

\section{Exemplary solution: Model of  $(H_2)_n$ chain }
\label{sec:model}

\begin{figure}
 \includegraphics[width=.5\textwidth]{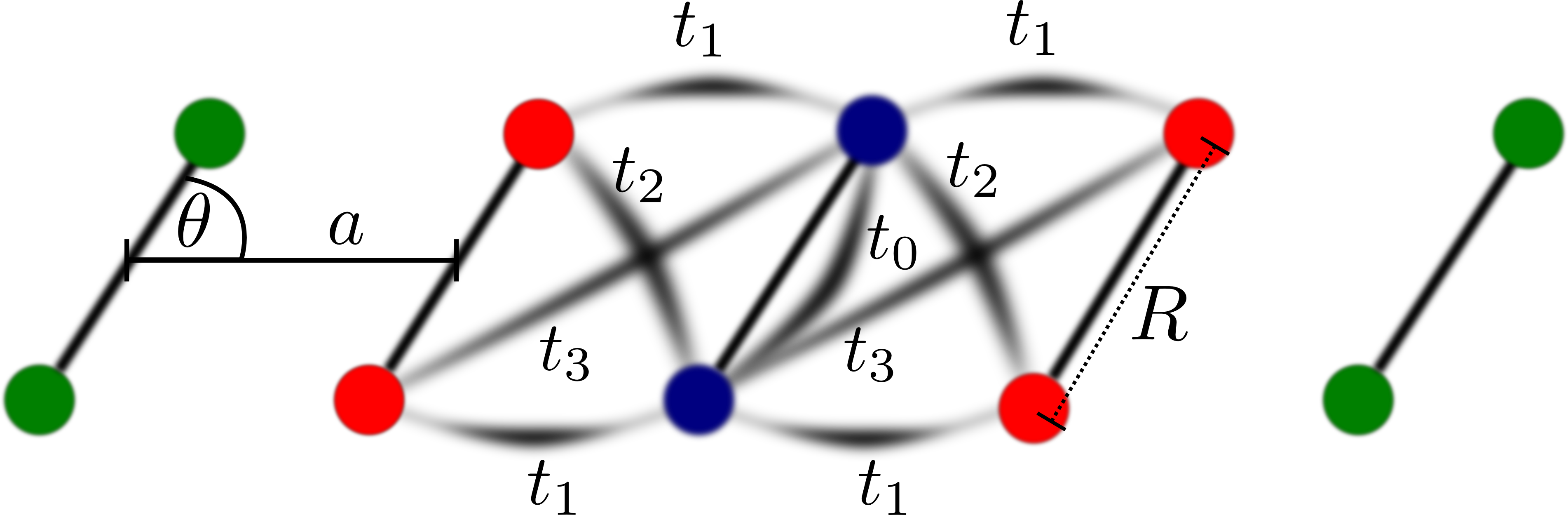}
 \caption{$(H_2)_n$ molecular chain, its parametrization with possible hoppings from/to central (blue) molecule. Green molecules are included as
 \emph{the background field} related to the system by the periodic boundary conditions.}
 \label{Fig:chain_model}
\end{figure}

We test our solution by means of  analysis performed for the chain consisting of $n=3$ hydrogen molecules stacked at
intermolecular distance $a$, with the molecule bond-length $R$, and the tilt angle $\theta$.
We regard this configuration as a part of periodic system  (cf. Fig.~\ref{Fig:chain_model}).
Hydrogen molecular chains  are interesting in view of the crucial role of electronic correlations in the molecules and related low-dimension systems \cite{Kadzielawa2, Spalek1, RycerzPhD}. 
The stability of the hydrogen molecular system was studied by means of variety of methods, e.g. DFT \cite{HydrogenClusters} or Self-Consistent Field (SCF) \cite{Kochanski,Kochanski2}, also in the 
context of the  existence of superfluidity \cite{Superfluidity}.

\begin{figure}
 \includegraphics[width=.5\textwidth]{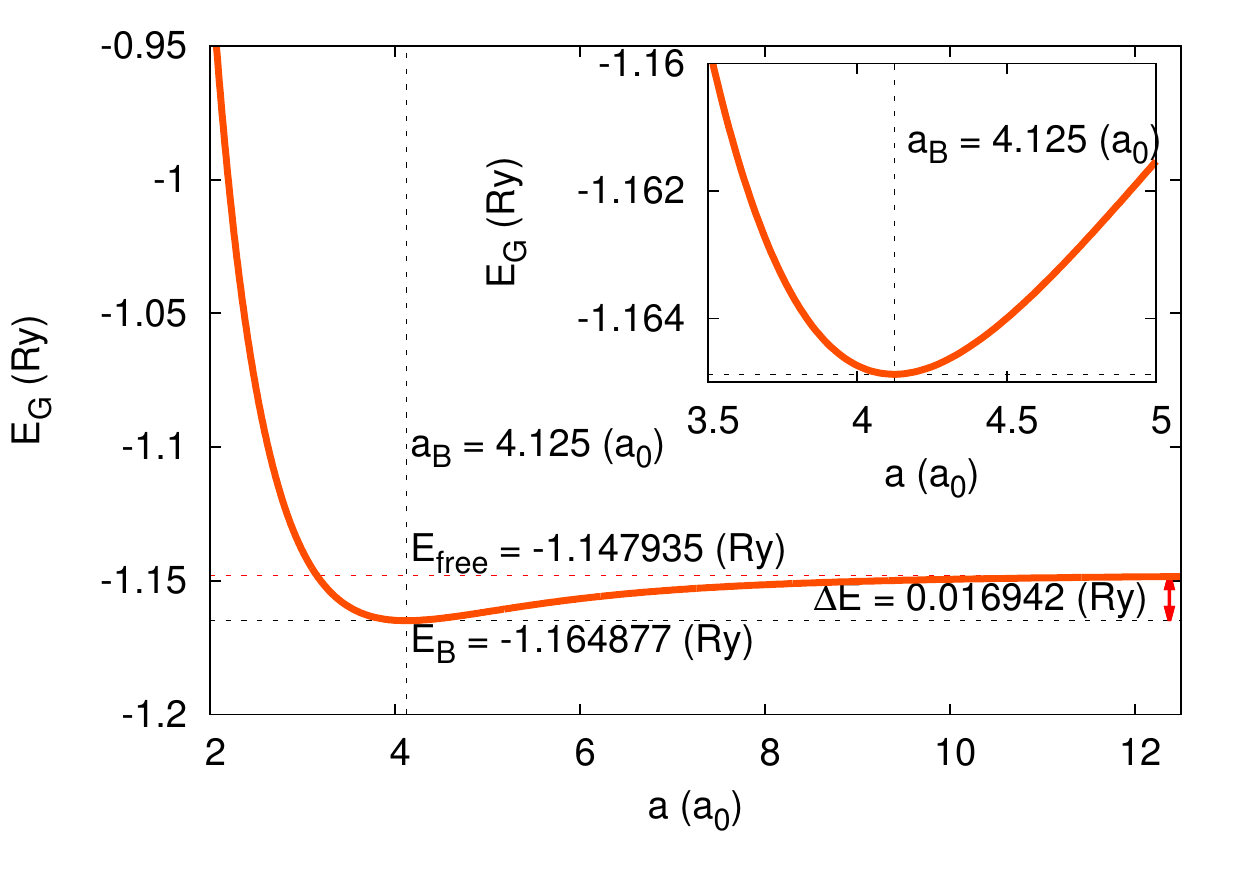}
 \caption{Ground state energy (per atom) for $(H_2)_3$ molecular chain as a function of the intermolecular distance $a$. Note the convergence to analytical solution \cite{Kadzielawa2} for the
 separate free molecule limit when $a \rightarrow \infty$.}
  \label{Fig:energy_vs_R}
\end{figure}
\begin{figure}
 \includegraphics[width=.5\textwidth]{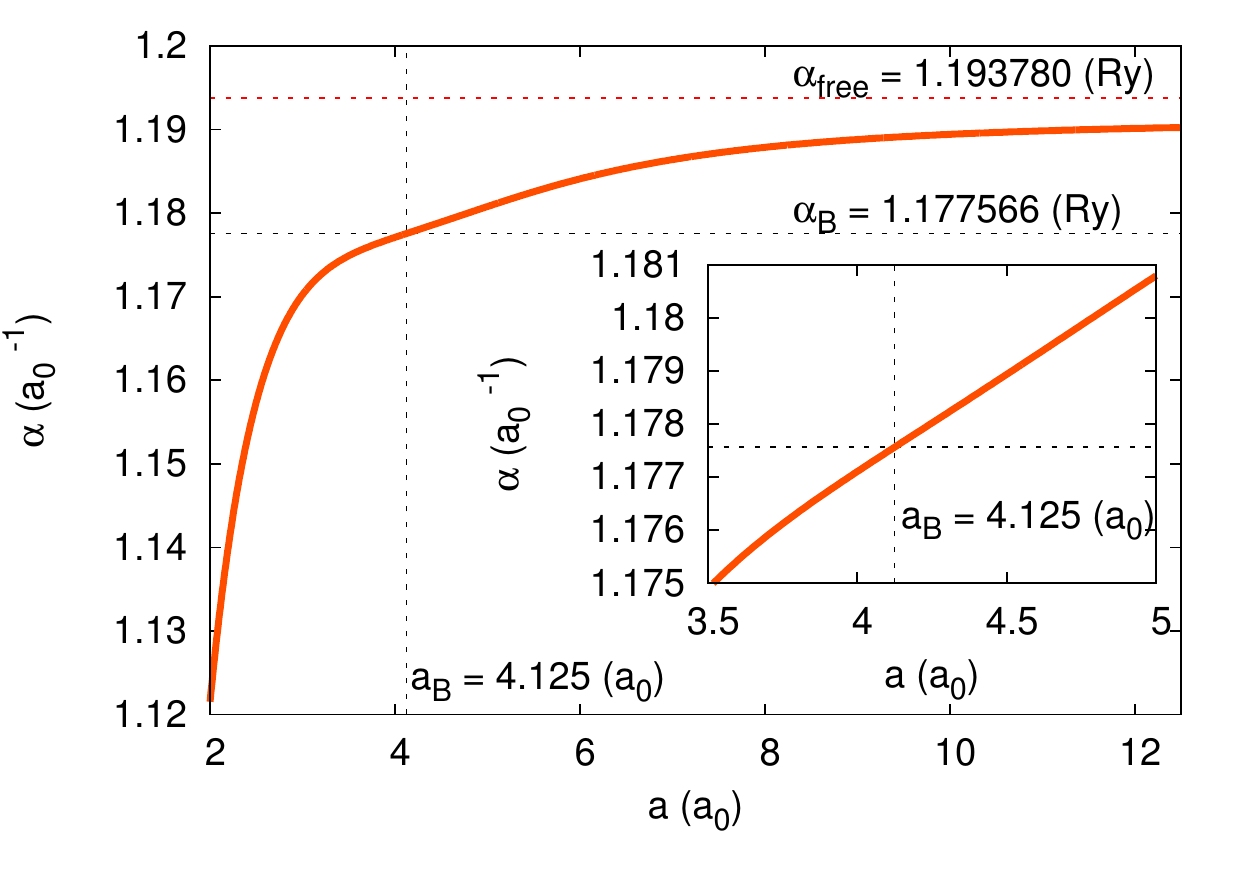}
  \caption{Inverse atomic orbital size $\alpha$ versus the intermolecular distance $a$. Note the convergence to the analytical solution \cite{Kadzielawa2} for
  the separate free molecule limit when $a \rightarrow \infty$.}
  \label{Fig:alpha}
\end{figure}

\begin{widetext}
\begin{figure}[H]
 \includegraphics[width=\textwidth]{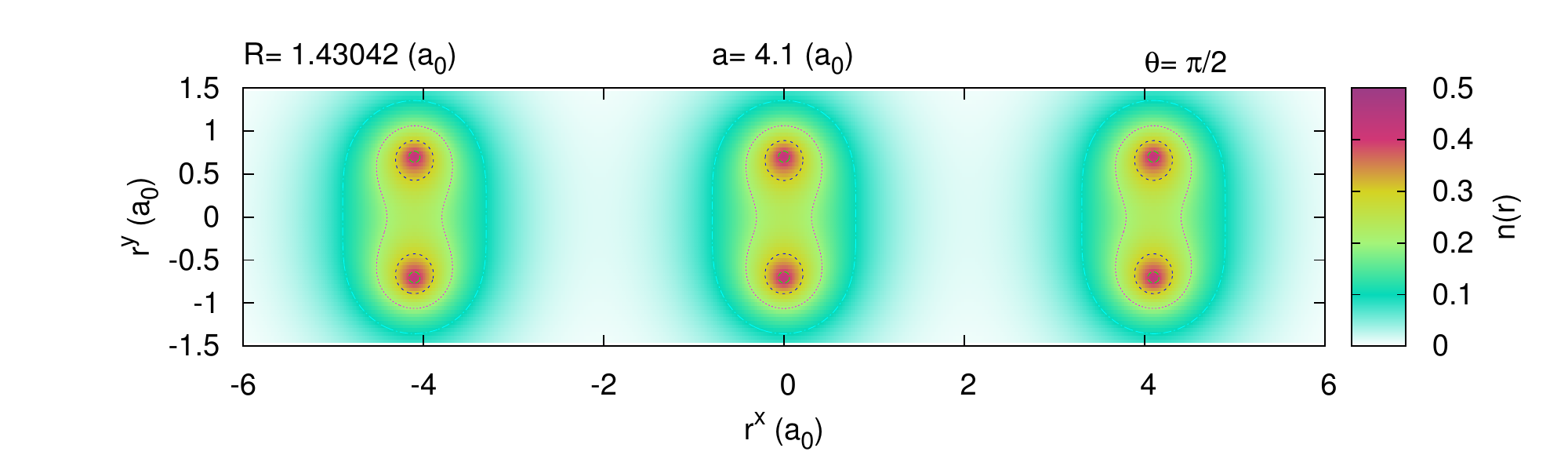}
 \caption{Electronic density $n(\vec{r})$ projected on $X-Y$ plane of the aligned $H_2$ molecules for the intermolecular distance close to the equilibrium value
 (marked by the vertical dashed line in Fig.~\ref{Fig:energy_vs_R}). The values of the structure parameters are specified.}
  \label{Fig:electron_density}
\end{figure}
\end{widetext}

We discuss the molecular hydrogen chain within the so-called \emph{extended Hubbard model}. This means that the interactions associated with the different atomic centers ($K_{ij}$)
are  taken into account. Eventually, the electronic part of Hamiltonian, with the ion-ion interaction explicitly included, can be rewritten then as composed of the three parts:
\begin{subequations}
\begin{align}
 \label{eq:h2n_hamiltonian_a}
  &\mathcal{H}_{Hubb} = \sum\limits_{i}\epsilon_i\NUM{i}{0} +\sum\limits_{ij\sigma}t_{ij}c_{i\sigma}^{\dagger}c_{j\sigma}^{}+U\sum\limits_{i}\NUM{i}{1}\NUM{i}{-1},\\
 \label{eq:h2n_hamiltonian_b}
  &\mathcal{H}_{ext} = \mathcal{H}_{Hubb} + \frac{1}{2}\sum\limits_{ij}K_{ij}\NUM{i}{0}\NUM{j}{0},\\
 \label{eq:h2n_hamiltonian_c}
  &\mathcal{H}_{tot} = \mathcal{H}_{ext} + \frac{1}{2}\sum\limits_{ij}\frac{2}{|\vec{R_i} - \vec{R_j}|} = \mathcal{H}_{ext} + \mathcal{V}_{i-i},
\end{align}
\end{subequations}
where the first two terms in \eqref{eq:h2n_hamiltonian_a} represent the single-particle energy with all possible hoppings $t_{ij}$
(to the fourth nearest neighbor, see Fig.~\ref{Fig:chain_model}) and are calculated with respect to \emph{the background field} (see Sec.~\ref{ssec:convergence} for details),
$U$ is on-site Coulomb repulsion (Hubbard term), $K_{ij}$ are intersite Coulomb repulsive interactions between supercell
and the \emph{background} sites (see \cite{RycerzPhD} and Appendix~\ref{app:Convergence} for details), $\mathcal{V}_{i-i}$ is the proton--proton repulsive interaction.
Despite its relative simplicity, the system exhibits non--trivial properties. However, as
we focus mainly on the computational aspects, we present here only the basic physical properties. Computational performance  tests of our solution are undertaken for arbitrarily
chosen  molecular bond--length $R=1.43042 (a_0)$, corresponding to the equilibrium value obtained by us previously for a single $H_2$ molecule  \cite{Kadzielawa2};
also $\theta = {\pi}/{2}$. The total number of electrons is equal to the  number of atomic centres ($6$ for $n=3$).
We test $E_G$ against the varying intermolecular distance $a$ (as shown in Fig.~\ref{Fig:energy_vs_R})  obtaining the van der Waals-like behavior of the total energy,
as expected \cite{Spalek1,RycerzPhD, Kochanski, Kochanski2}.
The single (spatially separated) $H_2$-molecule ground-state energy is  reproduced asymptotically for $a \rightarrow \infty$, as
marked in Fig.~\ref{Fig:energy_vs_R}. For the sake of completeness, we present in Fig.~\ref{Fig:alpha} the inverse atomic orbital size $\alpha$.
In Fig.~\ref{Fig:electron_density} we plot the contours of the electronic density $n(\vec{r})$ 
as the cross--section on the $X-Y$ plane close to the configuration related to the minimal value of $E_G$.
A very important feature of this solution worth mentioning is that as $a$ diminishes and approaches $R$ we observe a discontinuous phase transition from the molecular
to the atomic states, but this feature of the results are discussed elsewhere \cite{Kadzielawa3}. Also, the minimum energy provides a stable configuration
against the dissociation into separate molecules (cf. Fig.~\ref{Fig:energy_vs_R}).

\subsection{Convergence study}
\label{ssec:convergence}
\begin{figure}
  \includegraphics[width=.5\textwidth]{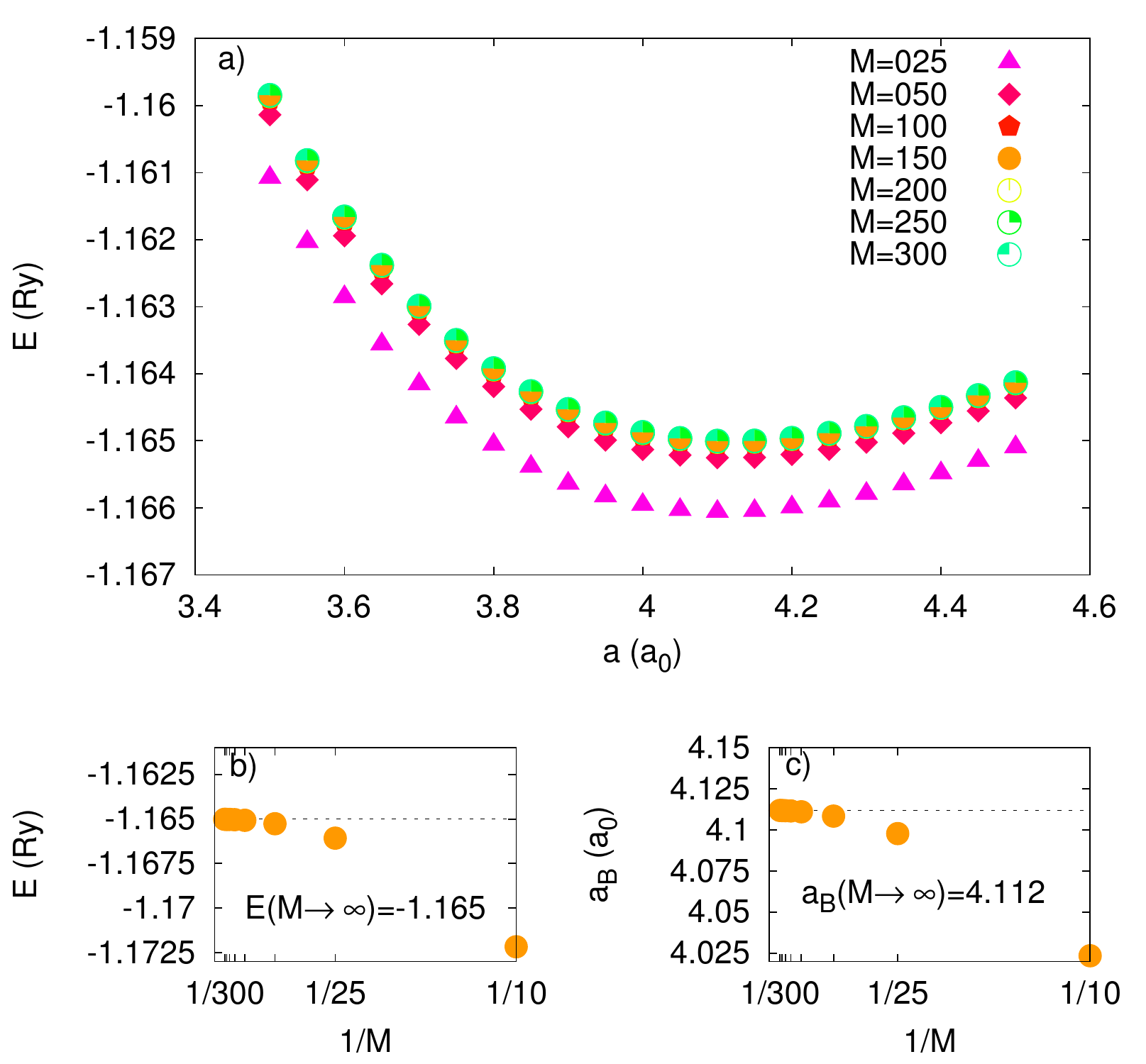}
  \caption{Convergence study: a)~ground-state energy $E_G$ versus intermolecular parameter $a$ close to the minimum for the different background field size $M$;
  b)~limit of $E_G$ for $M\rightarrow \infty$; c)~limit of the optimal intermolecular parameter $a_B$ (minimizing $E_G$) for $M\rightarrow \infty$. 
  The plots in b) and c) represent the finite-size scaling analysis.}
  \label{Fig:convergence_vs_background}
\end{figure}

\begin{figure}
  \includegraphics[width=.5\textwidth]{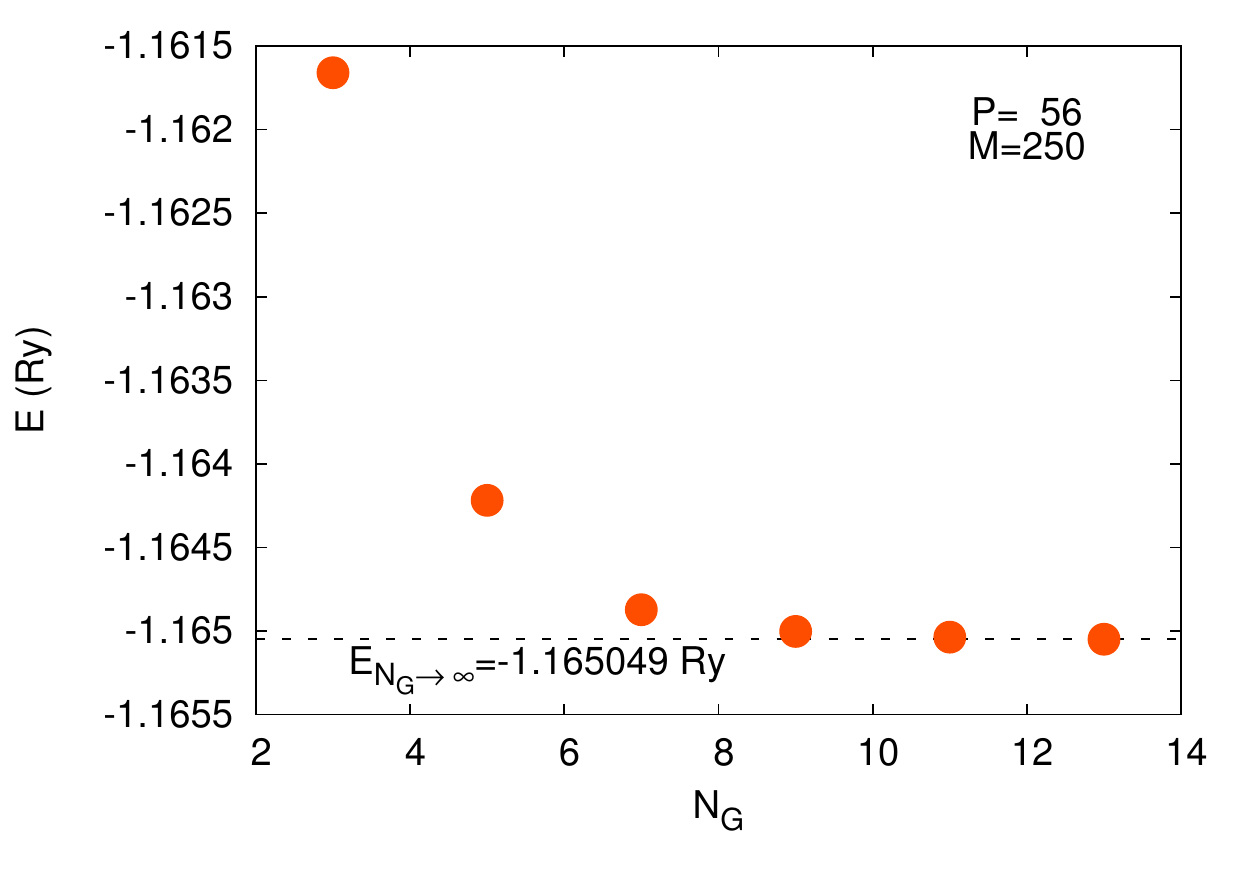}
  \caption{Convergence study: Ground-state energy $E_G$ versus number of Gaussians $N_G$ taken to represent the atomic basis and
  for the background field size $M=250$ and the number of SMP nodes $P=56$. }
  \label{Fig:convergence_vs_gauss}
\end{figure}

We have performed the convergence study to obtain parameters suitable for the speed--up analysis of the implemented approach. There are two most important components  playing crucial role:
number of Gaussian functions $N_G$ taken in contraction \eqref{eq:slater_definition} and the size of periodic \emph{background--field} of the  super-cell. The former does not require additional comment; 
it is not the case for the latter. With the  periodic boundary conditions  being imposed, a proper cut--off distance for the  interactions must be also established. The atomic energy $\epsilon_0$ becomes in general lower when a larger number of ionic centers are taken into account (i.e.,
the larger cut--off distance), then the contribution from \emph{the electron--ion} becomes stronger. On the other hand, analogical in nature but opposite in sign effects originate from the Coulomb \emph{ion--ion}  and  \emph{the electron--electron} repulsions.
In the limit $r_{cutoff}\to\infty$  both contributions cancel each other, as discussed in the context of EDABI in \cite{KurzykPhd} and, for the sake of completeness, in Appendix \ref{app:Convergence}.
One may note that the just described behavior is similar to the cancellation effect observed in  \emph{the jellium model} \cite{Giuliani}. We define $M$ as cut--off parameter, describing the size of background field as:
\begin{align}
 \label{eq:background_definition}
 &M = \frac{r_{cutoff}}{a}
\end{align}
In Fig.~\ref{Fig:convergence_vs_background} we plot the system total energy as a function of intermolecular distance, close to the energy-minimum configuration. It is clear that
if smaller cut--off distance is selected, the energy is underestimated. If cut--off was chosen to be $\geq150 a$, the consecutive
energies differ by less than the assumed numerical error in Lanczos matrix  diagonalization procedure (i.e. $10^{-4} Ry \sim 1 meV$). Therefore,
further calculations were carried out for $M=250$, what results in $510$ integrals to be calculated after
reductions caused by the system symmetries. As it follows from Fig.~\ref{Fig:convergence_vs_gauss}, $N_G = 9$ is the number of Gaussians, when the energy becomes convergent,
therefore the subsequent analysis  corresponds to  $N_G=9$ and $M=250$.


\subsection{Strong scaling for MPI+openMP solution}
\begin{figure}
 \includegraphics[width=.5\textwidth]{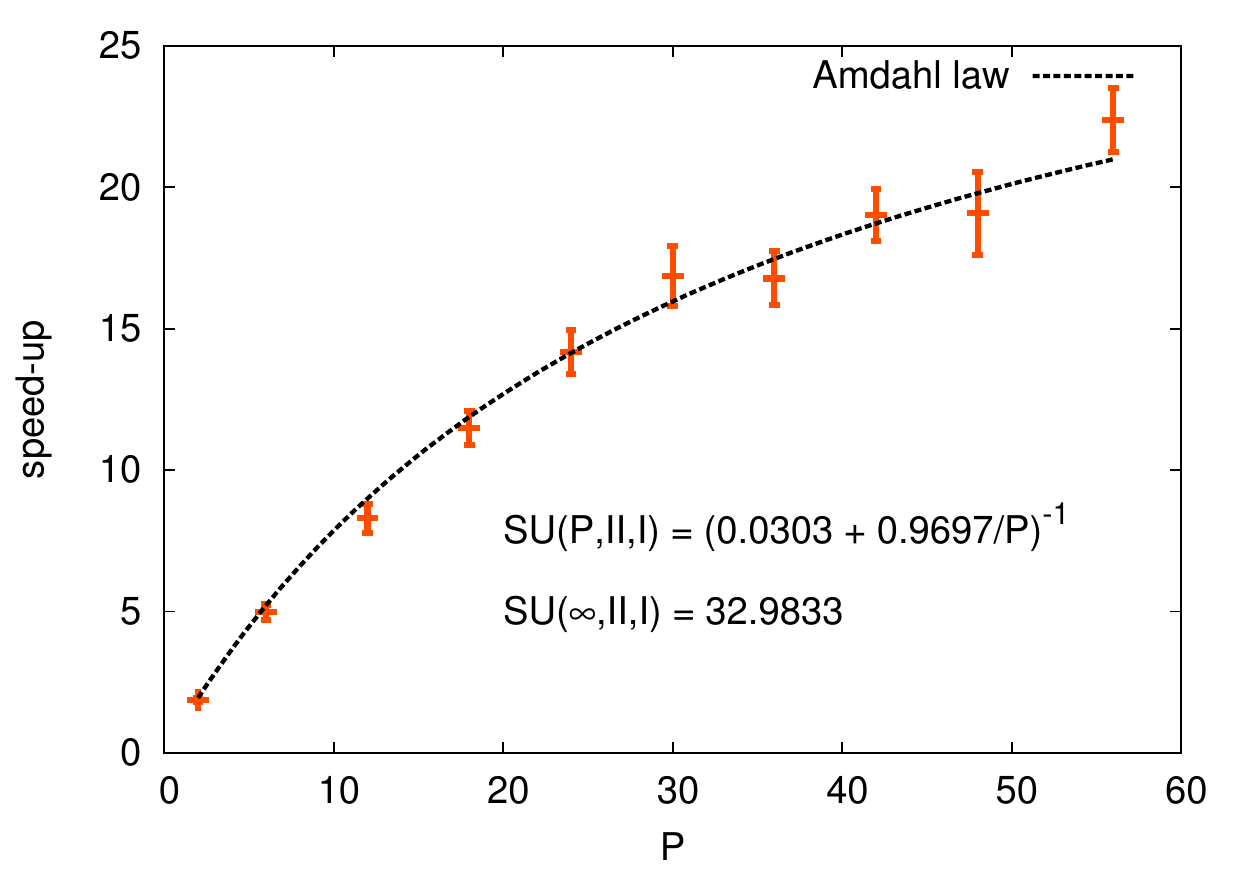}
 \caption{Speed-up ($SU$) as a function of number of processes $P$. Amdahl law curve fited to data. Each point probed 50 times.}
  \label{Fig:speed_up_mpi}
\end{figure}

Taking into account that we have one electron per atomic center in a two band system (molecule consists of two  hydrogen atoms), construction of the basis for 
three molecules leads to $924$ basis states. The diagonalization in terms of the iterative algorithm -- in this case Lanczos -- is not a challenge, 
especially if only the lowest eigenvalue is desired. A remark is necessary here: system described by substantially larger Hamiltonian matrices
can also be treated efficiently in the framework of the elaborated scheme. However, more sophisticated approaches (cf. Sec.~\ref{ssec:optimization_loop}) or approximate
methods are then indispensable. Therefore, the example we provide, fulfills the
requirement concerning the ratio of diagonalization to integration time. The latter is supposed to be the bottleneck during the execution of the optimization procedure.
We investigate the speed-up ($SU$) defined as
\begin{align}
 \label{eq:speed_up}
 &SU(P,X,Y) = \frac{T_Y(P=1)}{T_X(P)  },
\end{align}
with $T$ denotes time spent on the computation and
\begin{align}
 \label{eq:normalization}
 &X,Y \in \{I,II,\none \},
\end{align}
where first two symbols correspond  to  parallelism based on  the application of openMP and openMP+MPI respectively and \none \ is associated with sequential solution.
The calculations were performed for the  energy optimization at given ion configuration, close to the energy minimum, allowing to collect CPU time consumed by   $OP$.
The measurement were covered on HP server consisting of $96$ computational nodes each supplied with the two 8--cored Intel Xeon e5-2670 2.60GHz processors. 
The main board supported SMP  exposing one logical 16--core  processor per node thus $P$ refers to the number of SMP nodes.
Physically, the communication among nodes was provided by the InfiniBand 4X QDR interface. The  $SU(P,II,I)$ (Fig.~\ref{Fig:speed_up_mpi}) exhibits  Amdahl law--like behaviour \cite{Amdahl}. This law  states that speed--up limit 
in the strong--scaling regime can be described in terms of the following formula
 \begin{align}
  \label{eq:amdahl}
  &SU(P) = \frac{1}{1-f + \frac{f}{P}},
 \end{align}
where $f$ is the part of the program susceptible for the parallelization. The value of $f$   was found  by means of fit the Amdahl's law
to the obtained data. Approximate value of $f$  comes directly from the fit (see Fig.~\ref{Fig:speed_up_mpi} for details) and maximal $SU$ can also be estimated by
\begin{align}
 \label{eq:amdahls_limit}
 &SU_{max} \approx \lim_{P \to \infty}SU(P).
\end{align}
We found $f \approx 0.97$ and $SU_{max} \approx 33$, which confirms suitability of the application of our scheme. However, for the sake of completeness,
we investigated the Gaussian number threshold associated with the \emph{process--pool} solution efficiency. 

\subsection{Number of Gaussians and Efficiency Threshold}
\label{ssec:gaussians}

As mentioned above, the efficiency of the \emph{process-pool} solution depends on the workload assigned to each of the process in the pool. Notably, $N_G$ and
 the total number of integrals (associated with $M$ \eqref{eq:background_definition}) are the most important factors,
influencing robustness of the proposed approach.

\begin{figure}
 \includegraphics[width=.5\textwidth]{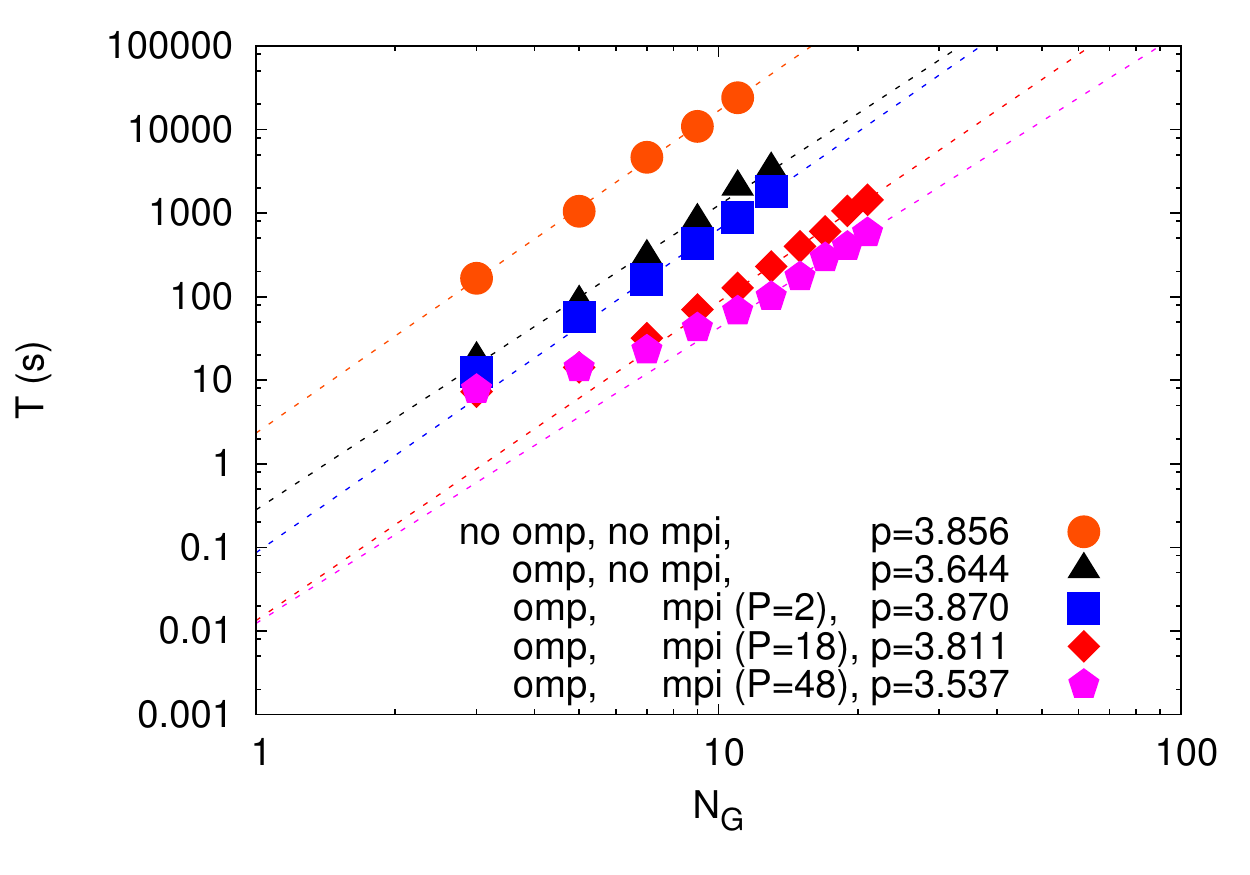}
 \caption{Computation time as a function of $N_G$ for sequential,
 only openMP, and MPI+openMP solutions. Note the linear behavior in the regime of large $N_G$ on log--log scale, suggesting $T \sim N_G ^p$. Values found from fit (see key)
 are consistent with ideal case $p=4$, where whole time is spent on calculating two-body integrals \eqref{eq:twobody_explicit}. As one-body integrals have $p=2$,
 values tend to drift from $4$.}
  \label{Fig:ng_time_coefs}
\end{figure}

We have performed measurements of computation time as a function of $N_G$ for a different number of $P$ (see Fig.~\ref{Fig:ng_time_coefs}).
For large number of Gaussians the optimization time has a universal scaling $T \sim N_G ^p$, with $p \approx 4$, meaning that
the two-particle integrals \eqref{eq:twobody_explicit} are the most computationally expensive, as expected.

\begin{figure}
 \includegraphics[width=.5\textwidth]{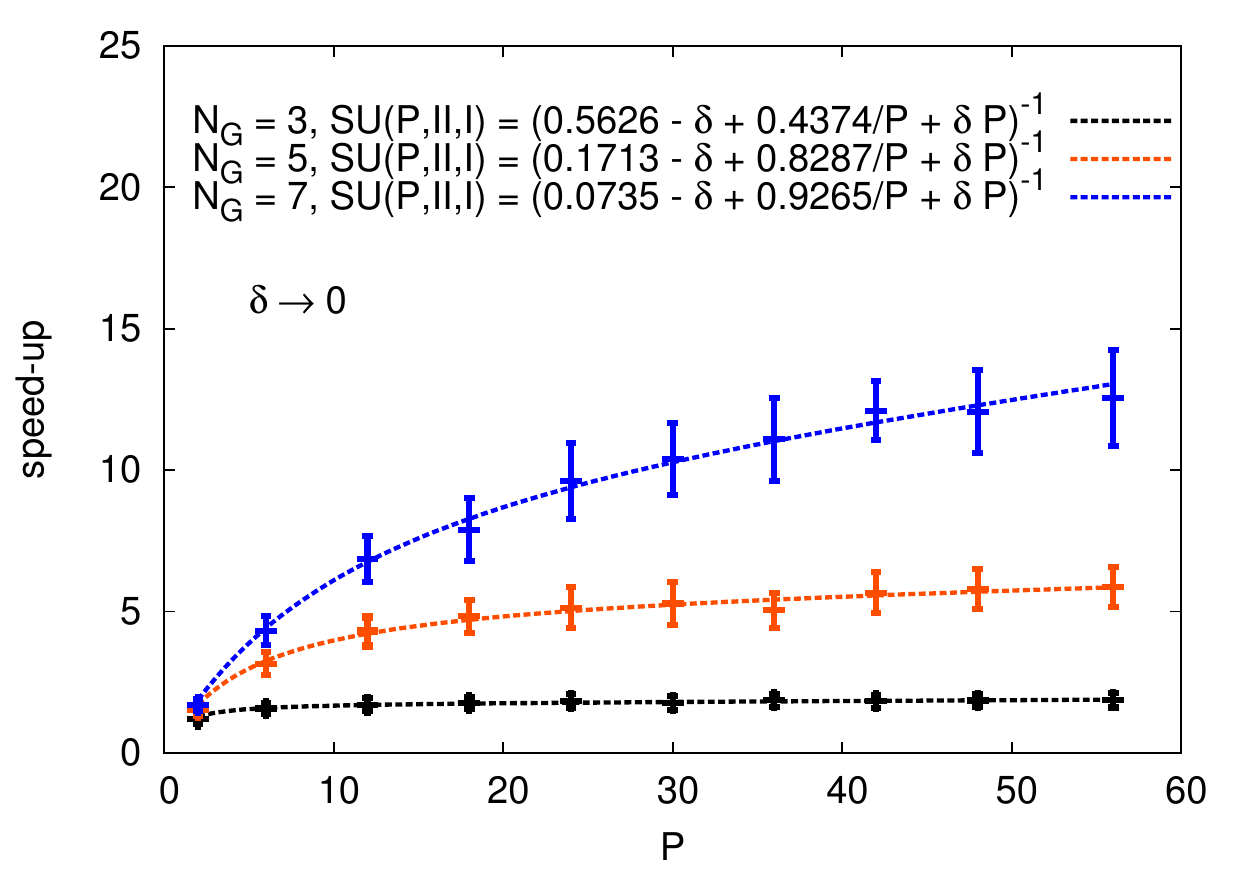}
 \caption{The relative speed-up $SU$ \eqref{eq:speed_up} versus number of nodes $P$ for $N_G \in \{3, 5, 7\}$ measured $100$ times. Note that when fitting the \emph{so-called} extended Amdahl law, communication overhead factor $\delta \approx 0$. Note that
 for $N_G=3$ only $44 \%$ of the program is performed parallelly.}
  \label{Fig:speed_up_diff}
\end{figure}

Following \cite{Schneider} we introduce the \emph{extended-Amdahl law} to include potential communication overhead. In our case,
the potentially most time-consuming (among $MPI$ communication routines used) $MPI \_ Gather$ routine scales lineraly with $P$. Taking this into account the speed--up  can be approximated 
by the following formula
\begin{align}
\label{eq:extended_Amdahl}
  SU_{comm}(P,Y,X) &= \frac{1}{1-f -\delta + \frac{f}{P} + \delta P}, \\\notag
\end{align}
where $\delta$ is constant to be determined. We performed fit of \eqref{eq:extended_Amdahl} to the speed--up as a function of $P$ for   $N_G \in \{3,5,7\}$ 
as we show in Fig.~\ref{Fig:speed_up_diff}. As expected $f$ decreases with decreasing $N_G$, but  $\delta < 10^{-4}$ even for $N_G=3$. This value is
negligible for the reasonable $P$ (the number of integrals is the upper bound) for any $f$. The lack of the communication overhead originates not only
from the utilization of \emph{InfiniBand} interface and linear scaling of the communication routine, but also from the small amount of data sent by each process to the \emph{root}
(e.g. $\sim 80$B for $P=50$). Hence the deviation from linearity for higher values of $P$  (see Fig.~\ref{Fig:ng_time_coefs}) comes from breaking the principal
assumption:

\noindent
For the lower numbers of Gaussians, the time of diagonalization (performed sequentially) becomes comparable or greater to that consumed
by the integrals computations. 

The analysis performed above allows to describe the boundaries where the proposed solution is effective. However, from the users perspective the most compelling feature is the absolute speed-up $SU(P,II,\none)$, as it is the metric for the time save. In the next paragraph we present this result.

\subsection{Absolute Speed-up}
Whilst analysis of the speed--up in strong scaling regime allowed us to investigate the efficiency gain in terms of a number of processes engaged in computation it is interesting and important to answer
what is the absolute acceleration  $SU(P,II,\none)$. Obviously, this quantity still depends on $P$. In Tab.~\ref{tab:SU} we show the values of the speed-up \eqref{eq:speed_up}
for different number of nodes $P$. The extremal case ($P=56$ with both levels of parallelization) the speed-up
\begin{align}
 \label{eq:TotalSpeedUp}
 SU \left(56, II, \none \right) = 303.418.
\end{align}
However, even for the lower number of SMP nodes ($P$) the absolute speed--up is excellent. For the sake of completeness, we retrieved the acceleration associated with openMP utilization (fourth column in Tab.~\ref{tab:SU}).
As each SMP node consisted of 16 cores the number of active threads was the same. We obtained  $SU(P,II,I)$ by means of the identity

\begin{align}
 \label{eq:su_omp}
 SU \left(P, II,I\right) = \frac{SU\left(P,II,\none\right)}{SU\left(P,I,\none\right)}.
\end{align}
We found good speed--up ratio $\sim 13$ (while the upper bound is $16$) coming from collapsing of nested loops \eqref{eq:twobody_explicit}.

\begin{table}[h]
\caption{Values of the speed-up (SU) for one-- and two-level parallelism for the different number of nodes $\mathbf{P}$.}
 \label{tab:SU}
\begin{tabular}{|l|l|l|l|}
\mc{1}{r|}{\textbf{P}} & \mc{1}{c|}{\textbf{SU(P,I,\none)}} & \mc{1}{c|}{\textbf{SU(P,II,\none)}} & \mc{1}{c}{\textbf{SU(P,II,I)}}\\\hline\hline
\mc{1}{r|}{2} & \mc{1}{d|}{ 1.866 } & \mc{1}{d|}{ 25.304 } & \mc{1}{d}{ 13.561 }\\\hline
\mc{1}{r|}{6} & \mc{1}{d|}{ 4.969 } & \mc{1}{d|}{ 67.382 } & \mc{1}{d}{ 13.561 }\\\hline
\mc{1}{r|}{12} & \mc{1}{d|}{ 8.288 } & \mc{1}{d|}{ 112.391 } & \mc{1}{d}{ 13.561 }\\\hline
\mc{1}{r|}{18} & \mc{1}{d|}{ 11.479 } & \mc{1}{d|}{ 155.656 } & \mc{1}{d}{ 13.561 }\\\hline
\mc{1}{r|}{24} & \mc{1}{d|}{ 14.174 } & \mc{1}{d|}{ 192.207 } & \mc{1}{d}{ 13.561 }\\\hline
\mc{1}{r|}{30} & \mc{1}{d|}{ 16.870 } & \mc{1}{d|}{ 228.763 } & \mc{1}{d}{ 13.561 }\\\hline
\mc{1}{r|}{36} & \mc{1}{d|}{ 16.779 } & \mc{1}{d|}{ 227.531 } & \mc{1}{d}{ 13.561 }\\\hline
\mc{1}{r|}{42} & \mc{1}{d|}{ 19.026 } & \mc{1}{d|}{ 258.007 } & \mc{1}{d}{ 13.561 }\\\hline
\mc{1}{r|}{48} & \mc{1}{d|}{ 19.085 } & \mc{1}{d|}{ 258.811 } & \mc{1}{d}{ 13.561 }\\\hline
\mc{1}{r|}{56} & \mc{1}{d|}{ 22.375 } & \mc{1}{d|}{ 303.418 } & \mc{1}{d}{ 13.561 }
\end{tabular}
\end{table}
\section{Conclusions}

We have presented an effective computational approach related to the EDABI method - quantum--mechanical approach allowing to treat the electronic correlations 
in a consistent manner. This means that we combine the second-quantization aspect (evaluation of energies for different many-particle configurations) with an explicit
evaluation of the renormalized wave functions (first-quantization aspect) in the resultant correlated many-particle state.
The number of microscopic parameters that are  necessary for description of  the physical system can be meaningful in many cases. Hence, their computations become challenging
as an effect of numerical complexity caused by the vast number of integrations to be performed \eqref{eq:microscopicP}.
Here we have addressed in detail the part of the whole problem that is associated with the single-particle basis optimization.
We have proposed the scheme based on the \emph{process--pool} concept enhanced by the \emph{two--level}
parallelism, and test it utilizing self--made generic implementation \cite{qmtURL} configured for the specific computational problem -- $\left(H_2\right)_n$ chain. 
The proposed approach is intuitive and has allowed us to speed up the  calculations significantly (of the order of $10^2$) while preserving its generic character.
Employing \emph{process--pool} solution to other systems is then straightforward.

Since the considered physical example serves  as an illustration of the elaborated scheme capability, one may consider engaging it to a wide class of computationally
advanced physical problems tractable within the framework of EDABI or similar methods. Such problems cover:
\begin{itemize}
 \item \textbf{lattice vibration (phonon) spectrum} via the so-called direct method (where all but few symmetries are lost, increasing the number of integrals dramatically -- from $510$ to over $10\ 000$
for $\left(H_2 \right)_n$ chain);
 \item calculation of \textbf{the electron--lattice coupling} parameters in direct space;
 \item electronic structure calculations of the \textbf{realistic two-- and three-dimensional} atomic and molecular crystals (e.g. hydrogen, lithium hydride), where both the number of atomic orbitals and \emph{the background field} increase essentially.
\end{itemize}

Neither form of the starting Hamiltonian nor the diagonalization scheme choice is essential for the applicability of the method, allowing us to incorporate other approaches 
such as the Gutzwiller approximation \cite{WysokinskiAbram, WysokinskiAbram1, Kadzielawa1} or Gutzwiller Wave Function - Diagramatic Expansion \cite{Kaczmarczyk} to
study molecular and extended spatially systems. In this manner, one can address e.g. the fundamental question of metallization in correlated systems \cite{Gebhard,Koch} with the explicit evaluation 
of an model parameters. So far we have been able to solve exactly the chain with $N=3$, $4$, $5$, and $6$ molecules and the results are of similar type \cite{Kadzielawa3}.
The finite-size type of scaling on the basis of these results requires additional analysis.

\section{Acknowledgments}
\label{sec:end}

The authors  are grateful to the Foundation for Polish Science (FNP) for financial support within the Project TEAM and to the National Science Center (NCN) for 
the support within the Grant MAESTRO, No.~DEC-2012/04/A/ST3/00342. We are also grateful to dr hab. Adam Rycerz from the Jagiellonian University for helpful comments.

\appendix
\section{EDABI method: A brief overview}
\label{app:EDABI}
The \textbf{E}xact \textbf{D}iagonalization \textbf{Ab} \textbf{I}nitio (EDABI) approach combines first- and second-quantization aspects when
solving the many-particle problem as expressed formally by its Hamiltonian.

The starting Hamiltonian \eqref{eq:hamiltonian} contains all possible dynamical processes starting from two-body interaction $V (r - r')$ in the coordinate
(Schr\"{o}dinger) representation. Its version \eqref{eq:hamiltonian_special} is already truncated and limited to two-state (here two-site) terms,
i.e., the three- and four-site terms have been neglected. The rationale behind this omission is that, as shown elsewhere \cite{Spalek1,Spalek2}, already
the two-site terms (matrix elements) $\sim J_{ij}$ and $V_{ij}$  are much smaller than those $\sim U$ and $K_{ij}$. Parenthetically, all the terms
are taken into account in the starting atomic basis composing the Wannier function. Nonetheless, there is no principal obstacle in including
all those terms in the case of the small systems considered here.

The second characteristic feature of EDABI is the single-particle basis optimization which composes the main topic of this paper.
If the basis defining the starting Wannier-function basis were complete (i.e., $L \rightarrow \infty$ in Eq.~\eqref{eq:wannier_definition}),
then no basis optimization is required and an exact solution is achieved. However, as our basis $\{ w_i \}$ is incomplete one,
in our view, we are forced to readjust the basis so that the system dynamics (correlations) are properly accounted for. This introduces a variational
aspect to our solution, since we introduce a variable wave function size, adjustable in the interacting (correlated) state.
This very feature represents one of the factors defining the method.
Such adjustment is also reasonable from a physical point of view, as the single-particle orbital adapts then itself to the presence of other particles (electrons).
In other words, the correlations induced by the predominant interaction terms ($\sim U$ and $K_{ij}$) influence the size and shape of the states
$\{ w_i (\vec{r} ) \}$. This means that the orbitals get renormalized in the process of the correlated state formation.
Nonetheless, it is not a priori determined that negligence of the higher virtually excited states in the expansion \eqref{eq:wannier_definition}
is minimized is such a manner. This should be tested and is one of the subjects of our current research. This is not
the primary topic of this paper so we shall not dwell upon it any further here.

Note also that by selecting the diagonalization of many-particle Hamiltonian in the second-quantized form, one avoids
writing the many-determinantal expansion of the multiparticle wave function, as is the case in the CI methods. However, out of our formulation
one can obtain the function $\Psi(\vec{r}_1,\dots,\vec{r}_N)$ and in particular, define many-particle covalency \cite{Spalek2}.
This transition from the Fock space back to the Hilbert space is possible as the two languages of description are equivalent
in the nonrelativistic situation (for a lucid and didactic exposition of the first- and second-quantization schemes and their equivalence
see e.g. \cite{Robertson}). The principal limitation of our method is the circumstance that it can be applied directly
only to relatively small systems when the exact diagonalization is utilized.

\section{Convergence of the single-particle energy for infinite system}
\label{app:Convergence}
In \eqref{eq:h2n_hamiltonian_a} we have the microscopic parameter contained in the single-particle energy expression, namely
\begin{align}
 \label{eq:eps_def}
 \epsilon_i = \matrixel{\psi_i}{\left( - \bigtriangledown^2 - \sum_{j} \frac{2}{\left| \vec{r} - \vec{R}_j \right|}\right)}{\psi_i},
\end{align}
where $i$ labels lattice site (node) at which we calculate this single-particle energy, and $j$ goes over the whole system.
Also, $\psi_i$ is the $1s$ Slater-type orbital centered on that site. For the sake of clarity, we disregard the orthogonalization procedure,
as the one-body parameters contain, strictly speaking, a linear combination of integrals in STO basis.

For an infinite system, the sum $\sum_{j}$ constitutes a series of the form \cite{Slater}
\begin{align}
\label{eq:eps_abinitio}
 \epsilon_i =& -\matrixel{\psi_i}{\bigtriangledown^2 }{\psi_i} - \sum_{j} \matrixel{\psi_i}{\frac{2}{\left| \vec{r} - \vec{R}_j \right|}}{\psi_i} \\\notag
	    =& \alpha ^2-2\alpha -\sum_{j} \frac{2}{\left| \vec{R}_{ij} \right|}+2\left(\alpha +\frac{1}{\left| \vec{R}_{ij} \right|}\right) e^{-2\alpha  \left| \vec{R}_{ij} \right|},
\end{align}
which diverges to $-\infty$.
In this Appendix we show that there is an \emph{effective} single-particle energy with no divergence for the infinite systems
(case more general than those discussed in \cite{RycerzPhD,KurzykPhd}).

We start from Hamiltonian \eqref{eq:h2n_hamiltonian_c}
\begin{align}
  &\mathcal{H}_{tot} = \mathcal{H}_{ext} + \mathcal{V}_{i-i},
\end{align}
where \emph{ion--ion} interaction is defined in the classical limit, i.e.,
 \begin{align}
 \label{eq:CRep}
  \mathcal{V}_{i-i} \equiv \frac{1}{2} \sum_{i \neq j} \frac{2}{\left| \vec{R}_{ij} \right|},
\end{align}
and $\left| \vec{R}_{ij} \right|$ is the distance between sites $i$ and $j$. We analyze next the remaining contributions, term by term.

\subsection{Intersite Coulomb term}
The intersite Coulomb term from \eqref{eq:h2n_hamiltonian_b} cab be rewritten in the form
 \begin{align}
 \label{eq:Ksplit}
 \notag \frac{1}{2}\sum\limits_{ij}K_{ij}\NUM{i}{0}\NUM{j}{0} = &\ComInEq{\frac{1}{2} \sum_{i \neq j} K_{ij} \delta\NUM{i}{0} \delta\NUM{j}{0}}{K^{(0)}} +  \ComInEq{\frac{1}{2} \sum_{i} \sum_{j} \left(1 - \delta_{ij} \right) K_{ij} \NUM{i}{0}}{K^{(1})} \\
				 + &\ComInEq{\frac{1}{2} \sum_{i} \sum_{j} \left(1 - \delta_{ij} \right) K_{ij} \NUM{j}{0}}{K^{(2)}} - \ComInEq{\frac{1}{2} \sum_{i \neq j} K_{ij}}{K^{(3)}}, 
 \end{align}
 where $\delta\NUM{i}{0} \equiv \left(\NUM{i}{0} - 1\right)$ and $\delta_{ij}$ is Kronecker's delta.
 
 We observe that when all sites are taken into account, the terms $K^{(1)}$ and $K^{(2)}$ are equivalent. We can rewrite them as follows
  \begin{align}
 \label{eq:Ksplit2}
 \mathcal{H}_{\text{K}} =& K^{(0)} - K^{(3)} +  2\ \frac{1}{2} \sum_{i} \sum_{j} \left(1 - \delta_{ij} \right) K_{ij} \NUM{i}{0} = \\\notag
                        \approx& K^{(0)} - K^{(3)} +  2\ \frac{1}{2} \sum_{i} \NUM{i}{0} \sum_{j(i)} K_{ij},
 \end{align}
 where $j(i)$ denotes the neighborhood of site $i$. Likewise,
 \begin{align}
  K^{(3)} =& \frac{1}{2} \sum_{i \neq j} K_{ij} = \frac{1}{2} \sum_{i} \sum_{j(i)} K_{ij} .
 \end{align}
 We can write finally that
  \begin{align}
 \label{eq:Ksplit3}
 \mathcal{H}_{\text{K}} =& K^{(0)} +  \frac{1}{2} \sum_{i} \NUM{i}{0} \sum_{j(i)}  K_{ij} \\\notag
                        +& \frac{1}{2} \sum_i \delta \NUM{i}{0} \sum_{j(i)} K_{{i}j}.
 \end{align}
 
 Note that for half-filling $\average{\NUM{i}{0}} = 1$ and the last part and $\average{K^{(0)}}$ disappears.
 
 \subsection{Ion-ion repulsion}
 Similarly, we can rewrite \eqref{eq:CRep} to the form
 \begin{align}
 \label{eq:Rsplit}
   \notag \mathcal{V}_{i-i} =&       \frac{1}{2} \sum_{i \neq j} \frac{2}{\left| \vec{R}_{ij} \right|} 
                             \approx \frac{1}{2} \sum_{i} \sum_{j(i)} \frac{2}{\left| \vec{R}_{ij} \right|} 
                            =       \frac{1}{2} N \sum_{j({i_0})} \frac{2}{\left| \vec{R}_{{i_0}j} \right|} \\
                            =&      \frac{1}{2} \sum_i \NUM{i}{0} \sum_{j({i})} \frac{2}{\left| \vec{R}_{{i}j} \right|}
                            -       \frac{1}{2} \sum_i \delta \NUM{i}{0} \sum_{j({i})} \frac{2}{\left| \vec{R}_{{i}j} \right|}.
 \end{align}
Again, the average of the latter term disappears for one particle per site.

\subsection{Total Hamiltonian}
We rearrange \eqref{eq:h2n_hamiltonian_c} obtaining so that the new form of Hamiltonian is
\begin{align}
 \label{eq:newHam}
  \mathcal{H_N} = \mathcal{H}_{\epsilon^\text{eff}} + \mathcal{H}_{\text{Hubbard}} + \mathcal{H}_{\delta n} .
\end{align}

The new terms are
\begin{align}
 \label{eq:neweps}
  \mathcal{H}_{\epsilon^\text{eff}} =  \sum_i \epsilon^{\text{eff}}_i \NUM{i}{0}, 
\end{align}
with $\epsilon^{\text{eff}}_i = \epsilon_i + 1/2 \sum_{j({i})} \left( {2}/{\left| \vec{R}_{{i}j} \right|} + K_{{i}j} \right)$,
\begin{align}
 \label{eq:newHubb}
  \mathcal{H}_{\text{Hubbard}} =& \sum\limits_{ij}^{}\sum\limits_{\sigma}t_{ij}c_{i\sigma}^{\dagger}c_{j\sigma}^{}+U\sum\limits_{i}\NUM{i}{1}\NUM{i}{-1},
  \end{align}
  \begin{align}
 \label{eq:deltan}
  \mathcal{H}_{\delta n} =& \frac{1}{2} \sum_{i \neq j} K_{ij} \delta\NUM{i}{0} \delta\NUM{j}{0} + \\\notag
  											&+ \frac{1}{2} \sum_i \delta \NUM{i}{0} \sum_{j(i)} \left(  K_{{i}j} - \frac{2}{\left| \vec{R}_{{i}j} \right|} \right).
\end{align}
$ $\\
$ $\\
$ $\\
The last question is whether the effective single-particle energy is now convergent.

\subsection{Convergence of the single-particle energy}
We can take $\epsilon^\text{eff}_i$ from \cref{eq:neweps,eq:eps_abinitio} and rearrange it in a following manner
\begin{align}
 \epsilon^{\text{eff}}_i =& \epsilon_i + \frac{1}{2} \sum_{j} \left( \frac{2}{\left| \vec{R}_{ij} \right|} + K_{ij} \right) \\\notag
			 =& \alpha ^2-2\alpha +\sum_{j} -\frac{2}{\left| \vec{R}_{ij} \right|}+2\left(\alpha +\frac{1}{\left| \vec{R}_{ij} \right|}\right) e^{-2\alpha  \left| \vec{R}_{ij} \right|} \\\notag
			  &+ \frac{1}{2} \sum_{j} \left( \frac{2}{\left| \vec{R}_{ij} \right|} + K_{ij} \right) \\\notag
			 =& \alpha ^2-2\alpha +\sum_{j} 2\left(\alpha +\frac{1}{\left| \vec{R}_{ij} \right|}\right) e^{-2\alpha  \left| \vec{R}_{ij} \right|} \\\notag
			  &+ \sum_{j} \left( \frac{1}{\left| \vec{R}_{ij} \right|} + \frac{1}{2}K_{ij} -\frac{2}{\left| \vec{R}_{ij} \right|}\right).
\end{align}
The latter part disappears in the classical limit $\left| \vec{R}_{ij} \right| \gg \alpha^{-1}$, where
\begin{align}
 K_{ij} \rightarrow \frac{2}{\left| \vec{R}_{ij} \right|},
 \end{align}
and the remaining part $\sum_{j} 2\left(\alpha +{\left| \vec{R}_{ij} \right| ^{-1}}\right) e^{-2\alpha  \left| \vec{R}_{ij} \right|}$
is convergent.

\bibliography{bibliography}
\end{document}